\documentclass[sigconf]{acmart}

\usepackage{amsmath}
\usepackage{mathrsfs}
\usepackage[linesnumbered,ruled,vlined]{algorithm2e}
\usepackage{booktabs}
\usepackage{multirow}
\usepackage{makecell}
\usepackage{float}
\usepackage{afterpage}
\usepackage{lipsum}
\usepackage{arydshln}
\usepackage{subfigure}
\usepackage{stfloats}
\usepackage{tcolorbox}
\usepackage{bbm}
\usepackage{balance}
\usepackage{titlesec}
\titlespacing*{\section}{0pt}{*1}{*0.5}
\titlespacing*{\subsection}{0pt}{*0.75}{*0.25}
\titlespacing*{\subsubsection}{0pt}{*0.75}{*0.25}

\definecolor{salmon}{HTML}{f8bec0}
\newtcbox{\besthighlight}[1][salmon]{
  on line, arc=0.75mm, colback=#1, colframe=#1, boxrule=0pt, boxsep=0pt, left=1.5pt, right=1.5pt, top=1pt, bottom=1pt
}

\AtBeginDocument{
  \providecommand\BibTeX{{
    \normalfont B\kern-0.5em{\scshape i\kern-0.25em b}\kern-0.8em\TeX}}}

\setcopyright{acmlicensed}
\copyrightyear{2025}
\acmYear{2025}

\acmConference[CIKM '25] {Proceedings of the 34th ACM International Conference on Information and Knowledge Management}{ November 10--14, 2025}{Seoul, Republic of Korea.}
\acmBooktitle{Proceedings of the 34th ACM International Conference on Information and Knowledge Management (CIKM '25), November 10--14, 2025, Seoul, Republic of Korea}

\acmISBN{979-8-4007-2040-6/2025/11}
\acmDOI{10.1145/XXXXXX.XXXXXX}

\begin{document}
\title{Harnessing Light for Cold-Start Recommendations: Leveraging Epistemic Uncertainty to Enhance Performance in User-Item Interactions}
\renewcommand{\shorttitle}{Leveraging Epistemic Uncertainty to Enhance Cold-Start Recommendations}

\author{Yang Xiang}
\orcid{0009-0007-9693-7963}
\authornote{Both authors contributed equally to this research.}
\affiliation{%
  \institution{Xi'an Jiaotong-Liverpool University}
  \city{Suzhou}
  \state{Jiangsu}
  \country{China}
}
\email{Yang.Xiang19@student.xjtlu.edu.cn}

\author{Li Fan}
\orcid{0009-0005-7234-5774}
\authornotemark[1]
\affiliation{%
  \institution{Xi'an Jiaotong-Liverpool University}
  \city{Suzhou}
  \state{Jiangsu}
  \country{China}
}
\email{Li.Fan21@student.xjtlu.edu.cn}

\author{Chenke Yin}
\orcid{0009-0009-2176-3967}
\affiliation{%
  \institution{Xi'an Jiaotong-Liverpool University}
  \city{Suzhou}
  \state{Jiangsu}
  \country{China}
}
\email{Chenke.Yin22@student.xjtlu.edu.cn}

\author{Menglin Kong}
\affiliation{%
  \institution{McGill University}
  \city{Montreal}
  \state{Quebec}
  \country{Canada}
}
\email{menglin.kong@mail.mcgill.ca}

\author{Chengtao Ji}
\orcid{0000-0001-5733-6881}
\authornote{Corresponding author.}
\affiliation{%
  \institution{Xi'an Jiaotong-Liverpool University}
  \city{Suzhou}
  \state{Jiangsu}
  \country{China}
}
\email{Chengtao.Ji@xjtlu.edu.cn}

\begin{abstract}
Most recent paradigms of generative model-based recommendation still face challenges related to the cold-start problem. Existing models addressing cold item recommendations mainly focus on acquiring more knowledge to enrich embeddings or model inputs. However, many models do not assess the efficiency with which they utilize the available training knowledge, leading to the extraction of significant knowledge that is not fully used, thus limiting improvements in cold-start performance. To address this, we introduce the concept of epistemic uncertainty to indirectly define how efficiently a model uses the training knowledge. Since epistemic uncertainty represents the reducible part of the total uncertainty, we can optimize the recommendation model further based on epistemic uncertainty to improve its performance. To this end, we propose a \textbf{C}old-Start \textbf{R}ecommendation based on \textbf{E}pistemic \textbf{U}ncertainty (CREU) framework. Additionally, CREU is inspired by Pairwise-Distance Estimators (PaiDEs) to efficiently and accurately measure epistemic uncertainty by evaluating the mutual information between model outputs and weights in high-dimensional spaces. The proposed method is evaluated through extensive offline experiments on public datasets, which further demonstrate the advantages and robustness of CREU.  The source code is available at \url{https://github.com/EsiksonX/CREU}.
\end{abstract}

\ccsdesc[500]{Information systems~Recommender systems}

\keywords{Recommender System, Cold-Start Recommendation, Uncertainty Quantification}

\maketitle

\section{Introduction}
In the fast-changing world of the digital information age, recommender systems (RecSys) have become essential tools for helping users find relevant content and items among the large amounts of information and options available. Despite their wide use, RecSys still faces ongoing issues, especially in "cold-start" situations, where there is little or no past interaction data for new users or items \cite{zhang2025cold,kong2024collaborative,hu2019hers}. In real-world situations, the cold-start problem can arise when new items are introduced, new users are added, or new platforms have limited interaction data. Recent approaches \cite{CVAR2022Zhao,hu2019hers,bai2023gorec,liu2023uncertainty,wang2024preference,xu2022alleviating} have been proposed to focus on improving the performance of RecSys in the cold-start setting through acquiring more knowledge to enrich embeddings or model inputs. GoRec \cite{bai2023gorec} alleviates the cold-start problem by generating latent representations for cold items based on multimedia content information. PAD-CLRec \cite{wang2024preference} addresses the cold-start problem by aligning content and collaborative features through a preference-aware dual contrastive learning approach. CVAR \cite{CVAR2022Zhao} generates enhanced warm-up item ID embeddings for cold items, leveraging both historical data and emerging interaction records.

Although these methods help mitigate the cold-start problem, they cannot determine whether increasing knowledge is effectively used. Epistemic uncertainty \cite{hullermeier2021aleatoric}, which refers to uncertainty arising from a lack of knowledge about the best model, stems from a model's ignorance and can be reduced with more data. Thus, using epistemic uncertainty can help us assess how the model utilizes knowledge and, based on this understanding, further reduce the cold-start problem. Previously, Monte Carlo (MC) methods have been used to estimate epistemic uncertainty due to the lack of closed-form solutions in most modeling scenarios \cite{depeweg2018decomposition,berry2023normalizing}. However, as the output dimension increases, these MC methods require a large number of samples to provide accurate estimates. The structure of PaiDEs \cite{berry2023escaping} offers a non-sample-based alternative for estimating information-based criteria in ensemble models with probabilistic outputs \cite{kolchinsky2017estimating}. However, PaiDEs estimate epistemic uncertainty using the Bhattacharyya distance or KL divergence. Since ensemble models are randomly initialized, the distribution differences between each component are significant (with little overlap). In this case, Sinkhorn divergence \cite{feydy2019interpolating} can better handle the differences or distance measurements between non-overlapping distributions.

Therefore, we propose a \textbf{C}old-Start \textbf{R}ecommendation based on \textbf{E}pistemic \textbf{U}ncertainty (CREU) framework to estimate epistemic uncertainty of ensemble models and further alleviate uncertainty based on it. The framework is divided into 3 phases: a) Ensemble \textbf{C}onditional \textbf{V}ariational \textbf{A}uto\textbf{E}ncoder (CVAE) \cite{sohn2015learning}, b) Epistemic Uncertainty based Optimization, c) Backbone Model. Specifically, we first introduce ensemble CVAEs, which not only use CVAEs but also apply an ensemble approach to provide input from different components for estimating epistemic uncertainty. Then, we use the structure of PaiDEs with the sinkhorn divergence to evaluate how consistent or inconsistent the component output distributions are. Finally, we use backbone models to predict results based on previous output. The main contributions of this work are summarized as follows:
\begin{itemize}
  \item We propose a method, CREU, to alleviate cold-start recommendation by taking advantage of epistemic uncertainty quantifying based on the pipeline of PaiDEs.
  \item Subsequently, we use Sinkhorn divergence to replace the Bhattacharyya distance or KL divergence, which are commonly used in PaiDEs due to the nearly non-overlapping nature of the distributions with different initializations.
  \item We conduct extensive experiments to demonstrate our proposed approach's effectiveness. We then assess the method's effectiveness and ignorance through ablation experiments to evaluate the validity of Epistemic Uncertainty-based Optimization. These experiments highlight the advantages of CREU over the state-of-the-art models.
\end{itemize} 

\section{Problem Formulation}  
Given a set of users $\mathcal{U}$ and items $\mathcal{I}$, each item is identified by its item ID $\mathbf{I}$. The features used for CTR prediction are denoted as $\mathcal{X}$, which can be categorical or continuous, such as user attributes and item characteristics. Moreover, we take part of the features as item side information $\mathcal{S} \subset \mathcal{X}$, which can be used to warm the item embeddings. We denote $\mathbf{v}_{\mathbf{I}} \in \mathbb{R}^{d}$, $\mathbf{v}_{\mathcal{X}} \in \mathbb{R}^{d \times |\mathcal{X}|}$, and $\mathbf{v}_{\mathcal{S}} \in \mathbb{R}^{d \times |\mathcal{S}|}$ as the embeddings of item $I$, features $X$, and side information $S$, respectively. The objective of CREU is to predict the CTR, denoted as $\hat{y} = f(f' (\mathbf{v}_{\mathbf{I}}, \mathbf{v}_{\mathcal{S}}), \mathbf{v}_{\mathcal{X}})$, where $f$ is the backbone model and $f'$ is the warm model. If $\hat{y}$ is close to 1, the user is more likely to click the item.

\section{Methodology}
\begin{figure*}[htbp]
  \centering
  \resizebox{0.6\textwidth}{!}{
    \includegraphics{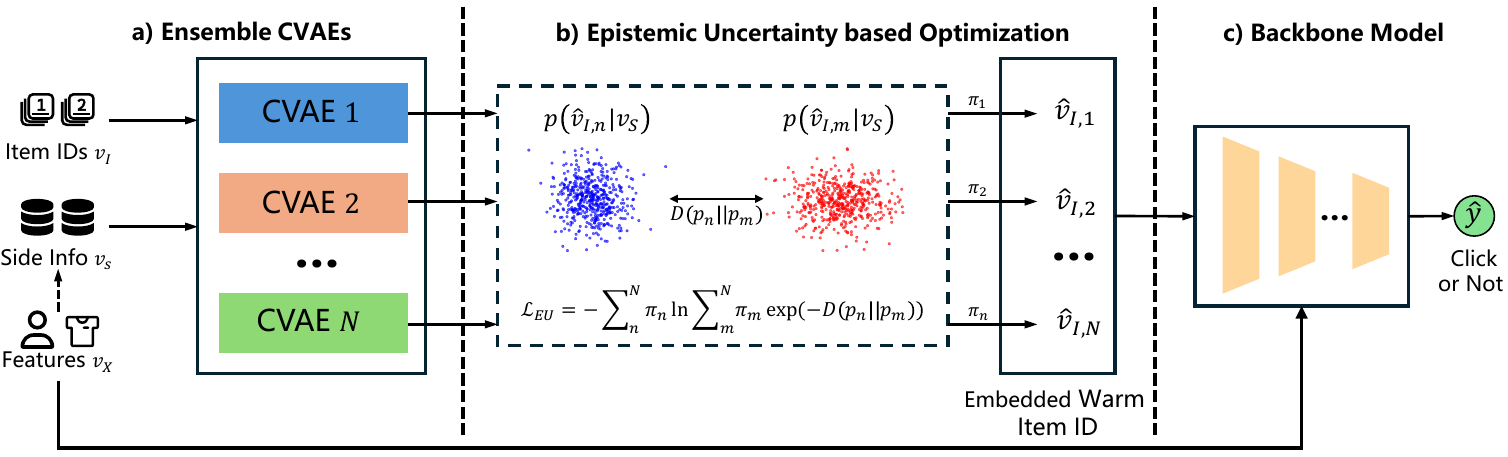}
  }
  \caption{The architecture of the proposed CREU model.}
  \label{fig:CREU}
  \vspace{-1.5em}
\end{figure*}

This section presents the CREU model, which aims to improve cold-start recommendations in CTR tasks \cite{CTR2007Richardson} by using epistemic uncertainty. CREU includes three components: (1) an ensemble of conditional variational autoencoders (CVAEs) for warming item embeddings, (2) a pairwise-distance estimator (PaiDE) optimization method to reduce epistemic uncertainty, thus enhancing training knowledge usage, and (3) a backbone model to predict CTR based on warmed embeddings and features. The CREU architecture is shown in Figure \ref{fig:CREU}.

\subsection{Ensembled CVAEs}
The ensemble of CVAEs aims to leverage side information to warm item embeddings. Our model uses $N$ number of CVAEs based on CVAR \cite{CVAR2022Zhao} to generate $N$ warmed embeddings for each item. The ensemble approach in our model leverages a form of randomization, allowing each to capture different perspectives of the training data. This diversity in learning helps to evaluate and reduce the model's ignorance. The CVAE can efficiently leverage side information to generate more informative embeddings. Note that all parameters of the CVAEs are randomly initialized, which means each CVAE can learn different aspects of the same input. For the $k$-th CVAE, where $k \in N$, the mapping of side information in the ensemble can be formulated as follows:
\begin{equation}
  \boldsymbol{\mu}_k, \boldsymbol{\sigma}_k = g_{\text{enc}, k}(\mathbf{v}_{\mathcal{S}}; w_{\text{enc}, k})
\end{equation}
\begin{equation}
  \mathbf{z}_k \sim \mathcal{N}(\boldsymbol{\mu}_k, \Sigma_k); \text{diag}(\Sigma_k) = \boldsymbol{\sigma}_k
\end{equation}
\begin{equation}
  \hat{\mathbf{v}}_{I,k} = g_{\text{dec}, k}(\mathbf{z}_k, x_{\text{freq}}; w_{\text{dec}, k})
\end{equation}
where $g_{\text{enc}, k}$ and $g_{\text{dec}, k}$ are the encoder and decoder functions, and $w_{\text{enc}, k}$ and $w_{\text{dec}, k}$ are their parameters. $x_{\text{freq}}$ is the frequency of item $i$ as a condition of the decoder to reduce the proportion of frequency bias. However, the distribution of the generated embeddings, derived from side information, differs from item embeddings based on item IDs, creating a distribution gap. To reduce this gap, we define a same encoder $g'_{\text{enc}, k}$ with parameters $w'_{\text{enc}, k}$ for item ID, and use $g_{\text{dec}, k}$ to generate embeddings $\hat{\mathbf{v}}_{I,k}'$ from item ID. We then use the Wasserstein distance to align the distribution of $\mathcal{N}(\boldsymbol{\mu}_k, \Sigma_k)$ to $\mathcal{N}(\boldsymbol{\mu}'_k, \Sigma_k')$:
\begin{equation}
  \mathcal{L}_w(w_{\text{enc},k}, w'_{\text{enc},k}, \pi_k) = \pi_k \cdot W(\mathcal{N}(\mu_k, \Sigma_k), \mathcal{N}_i(\mu_k', \Sigma_k'))
\end{equation}
where $\mu_k'$ and $\Sigma_k'$ are the mean and covariance of $\hat{\mathbf{v}}_{I,k}'$, $W$ is the Wasserstein distance, and $\pi_k$ is a differentiable weight for each CVAE. In the next section, we will use the ensemble CVAEs to evaluate whether the model fully utilizes the training knowledge.

\subsection{Epistemic Uncertainty-based Optimization}
\label{sec:eu}
The main challenge in the cold start problem is the lack of data, which leads to unstable training and suboptimal performance when the number of training epochs is limited \cite{10339320, zhu2020expert}. Epistemic uncertainty arises due to the model's lack of knowledge, often caused by insufficient or incomplete use of training data \cite{hora1996aleatory, der2009aleatory, hullermeier2021aleatoric}. Therefore, we aim to minimize the model's epistemic uncertainty to enhance its ability to utilize training data and ultimately improve the stability and performance under cold start conditions.

To capture the epistemic uncertainty, we first measure the full uncertainty of the model since uncertainty contains two components: epistemic and aleatoric. Following a supervised learning approach, we denote $\mathcal{D} = \{\mathbf{x}, \mathbf{y}\}$ as a dataset, where $\mathbf{x} \in \mathbb{R}^K$ is the input data and $\mathbf{y} \in \mathbb{R}^D$ is the corresponding label. To capture the uncertainty in the model, a common measure is the conditional differential entropy (CDE):
\begin{equation}
  H(y \mid x) = -\int p(y \mid x) \log p(y \mid x) dy.
\end{equation}
After we get the uncertainty of the distribution, the epistemic uncertainty can be expressed as described by \cite{houlsby2011bayesian}:
\begin{eqnarray}\label{eq:epistemic}
  I(y, \theta \mid x)=H(y \mid x)-E_{p(\theta)}[H(y \mid x, \theta)],
\end{eqnarray}
where $I(\cdot)$ refers to the mutual information and $\theta $ is the parameters of CVAEs, which $\theta \sim p(\theta)$. $H(y \mid x)$ represents total uncertainty estimated by CDE, and $E_{p(\theta)}[H(y \mid x, \theta)]$ reflects aleatoric uncertainty. Aleatoric uncertainty is caused by inherent noise or variability in the data, which cannot be reduced even with more data or better models.

However, $p(y | x)$ is complex and often exhibits multi-modal characteristics. Estimating epistemic uncertainty is challenging due to the high-dimensional space and the lack of closed-form solutions. Sampling approaches like the MC estimator require massive sampling to provide accurate estimates, which is computationally expensive \cite{depeweg2018decomposition}. To efficiently estimate epistemic uncertainty, we introduce Pairwise-distance Estimators (PaiDEs) \cite{berry2023escaping} to estimate epistemic uncertainty in our ensemble model. PaiDEs provide an efficient approach for estimating epistemic uncertainty by leveraging closed-form pairwise distances between the conditional probabilities of different components within the ensemble. This approach has been shown to effectively and accurately estimate epistemic uncertainty for high-dimensional continuous outputs \cite{berry2023escaping}. Denote that we have $N$ numbers of components in the ensemble model, and each component $m$ has a distribution $p_m=p_m(y | x, \theta_m)$. We can formulate PaiDEs to estimate epistemic uncertainty by generalized distance $D(p_m \| p_n)$ between two distributions $p_m$ and $p_n$ as follows:
\begin{eqnarray}\label{eq:paides}
  \begin{aligned}
    I(y, \theta \mid x) &:= H(y \mid x)-E_{p(\theta)}[H(y \mid x, \theta)] \\
    &=-\sum_{m=1}^N \pi_m \ln \sum_{n=1}^N \pi_n \exp \left(-D\left(p_m \| p_n\right)\right).
  \end{aligned}
\end{eqnarray}
Under our recommendation task, the epistemic uncertainty for an ensemble of CVAEs can be expressed as:
\begin{eqnarray}\label{eq:epistemic}
  \begin{aligned}
    I(\hat{\mathbf{v}}_{\mathbf{I}}, \theta \mid \mathbf{v}_{\mathcal{S}})=-\sum_{m=1}^N \pi_m \ln \sum_{n=1}^N \pi_n \exp \left(-D\left(\hat{\mathbf{v}}_{{\mathbf{I}},m} \| \hat{\mathbf{v}}_{{\mathbf{I}},n}\right)\right).
  \end{aligned}
\end{eqnarray}
In recommendation scenarios, the distributions of different components are often non-overlapping due to the sparse data in recommender systems \cite{DAMOUR2021644, chen2025data}. While in our setting, each CVAE is initialized with different parameters, which may exacerbate this phenomenon \cite{peng2025quantum, andersen2022evaluation}. Usually, the Kullback-Leibler (KL) divergence and Bhattacharyya distance are used as generalized distances in PaiDEs \cite{berry2023escaping, berry2024shedding}, these metrics are not well-suited for this non-overlapping distribution scenario. To address this issue, we employ Sinkhorn divergence \cite{feydy2019interpolating} as the generalized distance to measure non-overlapping distributions. Sinkhorn divergence, which incorporates entropy regularization, enables stable computation even when the support sets of the distributions do not overlap \cite{genevay2019sample}. The sinkhorn divergence is defined as:
\begin{eqnarray}\label{eq:sinkhorn}
  \begin{aligned}
    D(\hat{\mathbf{v}}_{{\mathbf{I}},m} \| \hat{\mathbf{v}}_{{\mathbf{I}},n})&:=\mathcal{W}_{\epsilon}(\hat{\mathbf{v}}_{{\mathbf{I}},m}, \hat{\mathbf{v}}_{{\mathbf{I}},n})\\
    &-\frac{1}{2}\left(\mathcal{W}_{\epsilon}(\hat{\mathbf{v}}_{{\mathbf{I}},n}, \hat{\mathbf{v}}_{{\mathbf{I}},n})+\mathcal{W}_{\epsilon}(\hat{\mathbf{v}}_{{\mathbf{I}},m}, \hat{\mathbf{v}}_{{\mathbf{I}},m})\right),
  \end{aligned}
\end{eqnarray}
where $\mathcal{W}_{\epsilon}(\cdot)$ is the regularized quadratic Wasserstein-2 distance \cite{mallasto2022entropy}, and $\epsilon$ is the entropy regularization parameter. By reducing the sinkhorn divergence-based epistemic uncertainty in our ensemble model, we can maximize the utility of the training data.

\subsection{CTR Prediction}
After estimating the epistemic uncertainty, we can use the warmed embeddings and features to predict the CTR. We use a backbone model $f$ to predict the CTR based on the warmed embeddings and features. The CTR prediction can be formulated as:
\begin{equation}
  \hat{y} = f(\hat{\mathbf{v}}_I, \mathbf{v}_{\mathcal{X}}; \theta),
\end{equation}
where $\hat{\mathbf{v}}_I = \sum_{k=1}^N \pi_k \cdot \hat{\mathbf{v}}_{I,k}$ is the ensemble of warmed embeddings, $\mathbf{v}_{\mathcal{X}}$ is the features, and $\theta$ are the parameters of the backbone model. Then, we can optimize the backbone model by minimizing the binary cross-entropy loss:
\begin{equation}
  \mathcal{L}_{\text{CTR}} = -\sum_{i=1}^N y_i \log \hat{y}_i + (1-y_i) \log (1-\hat{y}_i),
\end{equation}
where $y_i$ is the ground truth CTR, after training, our model can predict the CTR for recent emerging items based on its features and inferential warm embeddings.

\section{Experiments}
\subsection{Experiments Setup}
\subsubsection*{Datasets }
We evaluate the proposed CREU model on two public datasets: MovieLens-1M\footnote{\url{http://www.grouplens.org/datasets/movielens/}} and Taobao Display Ad Click\footnote{\url{https://tianchi.aliyun.com/dataset/56}}. To simulate cold start, items are ranked by total interactions—ratings for MovieLens and clicks for Taobao—and split into old (top $80\%$) and new (bottom $20\%$) items. New items are divided into four groups denoted as warm -a, -b, -c, and test set following the same setting as \cite{pan2019warm} and \cite{zhu2021learning}.

\subsubsection*{Backbones \& Baselines }
In the CTR prediction phase, we use the following respective backbones: DeepFM \cite{guo2017deepfm}, Wide\&Deep (W\&D) \cite{cheng2016wide}. To demonstrate the effectiveness of our proposed model, we choose some state-of-art methods for the item cold-start CTR prediction methods as baselines: DropoutNet \cite{volkovs2017dropoutnet}, Meta-E \cite{pan2019warm}, MWUF \cite{zhu2021learning}, and CVAR \cite{CVAR2022Zhao}. All hyperparameters are the same as the original papers or official repositories.

\subsubsection*{Implementation Details }
We implement the CREU model using PyTorch. The model is trained on a single NVIDIA RTX 4090 GPU. We use the Adam optimizer with a learning rate of $0.001$. The number of CVAEs is set to $3$, and the entropy regularization parameter $\epsilon$ is set to $0.15$. The number of iterations is set to $10$. The evaluation metrics are accuracy (ACC) and area under the curve (AUC).

\subsection{Performance Comparison}
We compare the performance of CREU with five state-of-the-art methods with backbones on the MovieLens-1M and Taobao Display Ad Click datasets. The results are shown in Table \ref{tab:performance}. CREU outperforms all baselines on both datasets in both cold and warm phases. For instance, on the MovieLens-1M dataset, CREU achieves an ACC of $74.36\%$ and an AUC of $63.85\%$ in the cold phase, which is $1.17\%$ and $1.54\%$ higher than the best baseline. In the warm phase, CREU achieves an ACC of $81.12\%$ and an AUC of $70.16\%$ in phase c, which is $0.63\%$ and $0.15\%$ higher than the best baseline. The results demonstrate the effectiveness of CREU in alleviating the cold-start problem and improving the performance of CTR prediction. Moreover, the performance of CREU in warm phase a has significantly improved compared to the best baseline, and it is close to the performance in warm phase c. This indicates that CREU can effectively utilize the training knowledge to improve the performance of cold-start recommendations.

\begin{table}
  \centering
  \caption{Performance comparison of CREU with state-of-the-art methods on the MovieLens-1M and Taobao Display Ad Click datasets. \besthighlight{Highlight} indicates the best performance.}
  \label{tab:performance}
  \vspace{-1em}
  \resizebox{0.8\linewidth}{!}{
    \begin{tabular}{c|c|cc|cc|cc|cc}
      \hline
      \multirow{2}{*}{Method} & \multirow{2}{*}{backbones} & \multicolumn{2}{c|}{Cold phase} & \multicolumn{2}{c|}{Warm phase a} & \multicolumn{2}{c|}{Warm phase b} & \multicolumn{2}{c}{Warm phase c}\\
      \cline{3-10}
      & & ACC & AUC & ACC & AUC & ACC & AUC & ACC & AUC\\
      \hline
      \multicolumn{10}{l}{\textbf{\textit{Dataset: MovieLens-1M}}}\\
      \hdashline
      DeepFM & & $72.67$ & $62.31$ & $74.23$ & $63.83$ & $75.74$ & $65.03$ & $76.94$ & $66.08$\\
      \hdashline
      DropoutNet & \multirow{5}{*}{DeepFM} & $73.87$ & $63.39$ & $74.91$ & $64.41$ & $75.87$ & $65.31$ & $76.73$ & $65.99$\\
      Meta-E & & $73.27$ & $63.44$ & $74.41$ & $64.32$ & $75.44$ & $65.19$ & $76.33$ & $65.92$\\
      MWUF & & $73.16$ & $62.89$ & $74.62$ & $64.13$ & $75.89$ & $65.21$ & $77.01$ & $66.16$\\
      CVAR & & $74.19$ & $63.56$ & $79.27$ & $67.89$ & $80.21$ & $68.56$ & $80.41$ & $69.78$\\
      CREU (Our) & & $\besthighlight{74.36}$ & $\besthighlight{63.85}$ & $\besthighlight{79.53}$ & $\besthighlight{67.98}$ & $\besthighlight{80.83}$ & $\besthighlight{69.21}$ & $\besthighlight{81.12}$ & $\besthighlight{70.16}$\\
      \hdashline
      W\&D & & $70.71$ & $59.72$ & $72.32$ & $61.64$ & $73.54$ & $62.73$ & $74.61$ & $63.72$\\
      \hdashline
      DropoutNet & \multirow{5}{*}{W\&D} & $71.25$ & $60.38$ & $72.28$ & $61.59$ & $73.13$ & $62.44$ & $73.90$ & $63.14$\\
      Meta-E & & $71.13$ & $60.41$ & $72.21$ & $61.54$ & $73.07$ & $62.39$ & $73.83$ & $63.09$\\
      MWUF & & $70.63$ & $59.66$ & $72.30$ & $61.57$ & $73.55$ & $62.75$ & $74.59$ & $63.66$\\
      CVAR & & $69.37$ & $56.43$ & $76.27$ & $65.25$ & $77.56$ & $66.39$ & $78.40$ & $67.21$\\
      CREU (Our) & & $\besthighlight{72.55}$ & $\besthighlight{60.89}$ & $\besthighlight{77.90}$ & $\besthighlight{67.23}$ & $\besthighlight{78.25}$ & $\besthighlight{66.63}$ & $\besthighlight{78.48}$ & $\besthighlight{67.34}$\\
      \hline
      \multicolumn{10}{l}{\textbf{\textit{Dataset: Taobao Display Ad Click}}}\\
      \hdashline
      DeepFM & & $59.83$ & $13.50$ & $60.97$ & $13.78$ & $62.07$ & $14.01$ & $63.11$ & $14.38$\\
      \hdashline
      DropoutNet & \multirow{5}{*}{DeepFM} & $59.89$ & $13.52$ & $60.98$ & $13.47$ & $62.03$ & $13.96$ & $63.02$ & $14.35$\\
      Meta-E & & $59.82$ & $13.46$ & $60.93$ & $13.77$ & $61.95$ & $14.00$ & $62.94$ & $14.28$\\
      MWUF & & $59.86$ & $13.48$ & $60.82$ & $13.74$ & $61.84$ & $13.99$ & $62.79$ & $14.29$\\
      CVAR & & $59.78$ & $13.47$ & $61.98$ & $14.80$ & $63.08$ & $14.77$ & $\besthighlight{63.80}$ & $15.03$\\
      CREU (Our) & & $\besthighlight{60.03}$ & $\besthighlight{13.53}$ & $\besthighlight{62.41}$ & $\besthighlight{14.83}$ & $\besthighlight{63.11}$ & $\besthighlight{14.81}$ & $63.73$ & $\besthighlight{15.05}$\\
      \hdashline
      W\&D & & $60.81$ & $13.60$ & $61.29$ & $14.27$ & $62.07$ & $14.55$ & $62.87$ & $14.84$\\
      \hdashline
      DropoutNet & \multirow{5}{*}{W\&D} & $\besthighlight{60.95}$ & $13.59$ & $61.84$ & $14.27$ & $62.46$ & $14.54$ & $63.12$ & $14.74$\\
      Meta-E & & $60.82$ & $13.78$ & $61.22$ & $14.43$ & $61.90$ & $14.77$ & $62.59$ & $15.06$\\
      MWUF & & $60.89$ & $13.82$ & $61.25$ & $14.23$ & $62.10$ & $14.57$ & $62.85$ & $14.83$\\
      CVAR & & $60.51$ & $13.68$ & $62.20$ & $14.57$ & $62.90$ & $14.95$ & $63.36$ & $15.11$\\
      CREU (Our) & & $60.79$ & $\besthighlight{13.79}$ & $\besthighlight{62.45}$ & $\besthighlight{14.58}$ & $\besthighlight{63.11}$ & $\besthighlight{15.00}$ & $\besthighlight{63.73}$ & $\besthighlight{15.15}$\\
      \hline
    \end{tabular}
  }
  \vspace{-2em}
\end{table}

\subsection{Ablation Study}
To evaluate the effectiveness of the proposed CREU model, we conduct an ablation study on the MovieLens-1M dataset. The study analyzes the contributions of each component in CREU by comparing the full CREU model against its ablated versions and analyzes the quantification of effectiveness through computing epistemic uncertainty in each iteration. 

\subsubsection{Effectiveness of Each Component}
The aim of this experiment is to measure the contribution of each component in CREU. According to the results in Figure~\ref{fig:ablation}, using an ensemble CVAE already outperforms the regular CVAE. However, when the ensemble CVAE is enhanced with epistemic uncertainty optimization, which is the approach used in CREU, further improvements in performance can be observed. In the cold phase, the CREU model shows smoother performance curves at each iteration compared with the other two models, indicating that epistemic uncertainty optimization makes better use of the training data to support the cold-start process. In other phases, CREU continues to achieve better results, suggesting that the other models do not fully utilize the additional knowledge and that the epistemic uncertainty optimization helps improve the use of training information.

\subsubsection{Utilization Analysis of Training Knowledge through Epistemic Uncertainty}
\begin{figure}[t]
  \centering
  \caption{Ablation study of CREU on the MovieLens-1M dataset. (1st row: DeepFM backbone, 2nd row: W\&D backbone, Color bar: epistemic uncertainty.)}
  \label{fig:ablation}
  \resizebox{0.8\linewidth}{!}{%
    \includegraphics{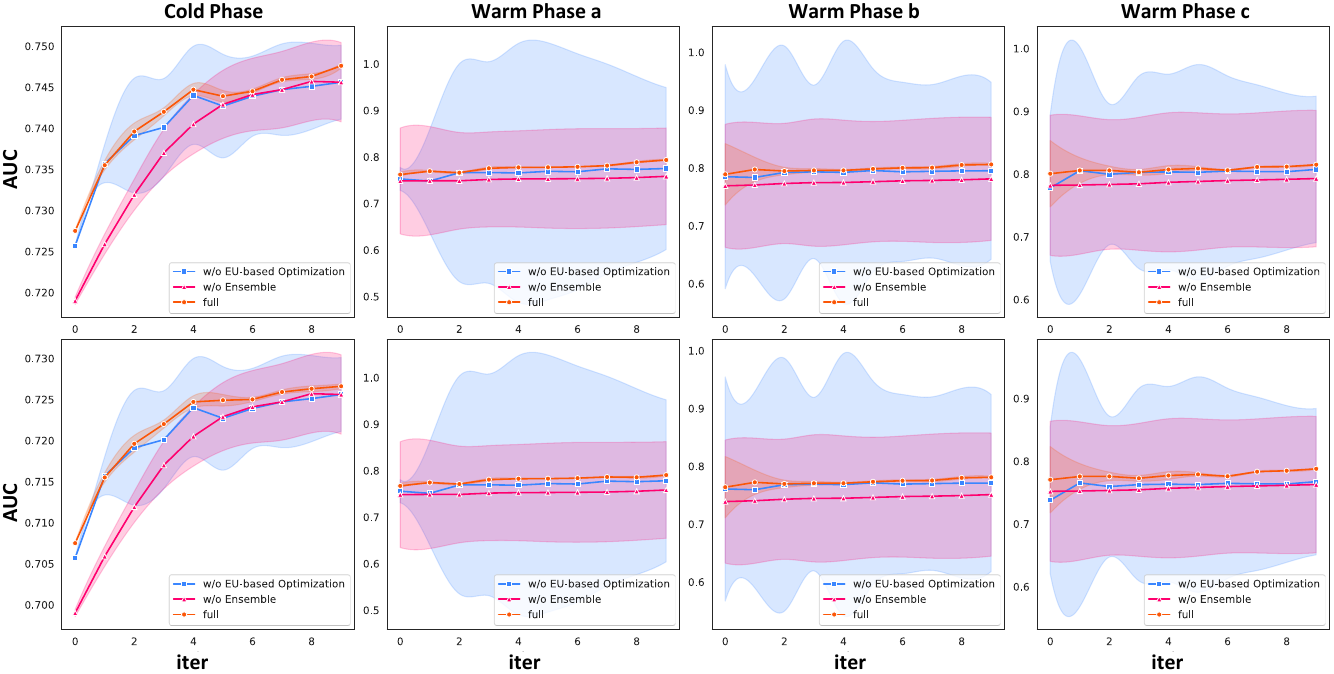}
  }
  \vspace{-2em}
\end{figure}

We calculated the epistemic uncertainty for three variants: the single CVAE, the Ensemble CVAE, and the CREU (i.e., the complete model) using Equation~\eqref{eq:epistemic}. Since the single CVAE does not produce multiple outputs, we run it several times to obtain multiple results and then compute its uncertainty. Figure~\ref{fig:ablation} displays the epistemic uncertainty for each iteration by showing the width of the corresponding curves. First, we observe that with epistemic uncertainty-based optimization, the uncertainty converges to a very low value by the final iterations. Second, the single CVAE shows no change in uncertainty, which agrees with our expectations and indicates that it does not fully utilize the extra information it extracts, leading to a performance gap with CREU. In contrast, the Ensemble CVAE exhibits large fluctuations because of the variance among multiple models; however, due to the ensemble properties, its uncertainty remains at a relatively high level. Overall, our visualization of uncertainty confirms that epistemic uncertainty reflects the degree of knowledge utilization, and that improving this utilization leads to better performance.

\section{Conclusion}
In this paper, we propose a novel model, CREU, to address the cold-start problem in CTR prediction tasks. CREU leverages an ensemble of CVAEs to generate warmed embeddings and an epistemic uncertainty-based optimization method to reduce the model's ignorance of the training data. Experimental results on two public datasets demonstrate that CREU outperforms state-of-the-art methods in both cold and warm phases. Furthermore, an ablation study shows that the ensemble CVAEs and epistemic uncertainty-based optimization are effective in improving the performance and reducing the ignorance of the model.

\begin{acks}
This work was partially supported by the Suzhou Municipal Key Laboratory for Intelligent Virtual Engineering (SZS2022004).
\end{acks}

\bibliographystyle{ACM-Reference-Format}
\balance
\bibliography{ref}

\end{document}